\documentclass[
  aps,
  prl,
  reprint,
  superscriptaddress,
  nofootinbib
]{revtex4-2}

\usepackage{amsmath,amssymb,bm}
\usepackage{graphicx}
\usepackage{dcolumn}
\usepackage{hyperref}
\usepackage{xcolor}
\usepackage{xurl}

\begin{document}

\title{
From Equilibrium Criticality to Universal Collective Phases through Nonreciprocal Coupling of Ordered Systems}

\author{Fabrizio Rippa}
\email{fabrizio.rippa@unicampania.it}
\affiliation{Department of Mathematics and Physics, University of Campania "Luigi Vanvitelli", Italy}

\author{Jacopo A. Garofalo}
\affiliation{Department of Mathematics and Physics, University of Campania "Luigi Vanvitelli", Italy}

\author{Eugenio Lippiello}
\affiliation{Department of Mathematics and Physics, University of Campania "Luigi Vanvitelli", Italy}

\date{\today}

\begin{abstract}
Nonreciprocal interactions have emerged as a fundamental driver of collective behavior far from equilibrium. Here, we show that nonreciprocal couplings between spin systems whose isolated dynamics favors ordered states, beyond the previously identified oscillatory (swap) phase, give rise to a richer nonequilibrium phenomenology, including a multistable regime where ordered and disordered states coexist, a scenario strictly forbidden in equilibrium. Crucially, we show that this nonequilibrium phenomenology is inherited directly from the critical properties of the underlying isolated model.  In particular, tricriticality in the isolated system gives rise to multistability, hysteresis, and hard excitations in the oscillatory dynamics, while non-normal interactions further enrich the phase diagram by generating multistability.  Moving beyond mean-field theory, we derive a universal field equation governing the onset of these nonreciprocal oscillations in finite dimensions.
Our results establish a unified theoretical framework connecting critical phenomena, dynamical systems, and non-reciprocal collective dynamics, with applications ranging from opinion dynamics to active matter.
\end{abstract}

\maketitle
Originally developed to describe equilibrium magnetism \cite{brush1967history,wu1982potts}, spin models have become paradigmatic frameworks for collective phenomena in interacting systems, including nonequilibrium dynamics, critical behavior on complex networks, and social dynamics \cite{schmittmann1995statistical,dorogovtsev2008critical,castellano2009statistical}. In
these models, the state of each unit is represented by a binary spin
$s_i=\pm1$, while the stochastic transition rate
$W(s_i\rightarrow -s_i)$ determines the microscopic dynamics.  When the
transition rates satisfy detailed balance, the dynamics relaxes toward
the Boltzmann distribution associated with a Hamiltonian
$\mathcal H$. This framework includes the Ising model and its
extensions, such as the random-field Ising model (RFIM), whose
equilibrium critical behavior is well understood. By contrast, several
nonequilibrium models, including the $q$-voter model~\cite{castellano2009nonlinear,nyczka2012phase}, violate detailed
balance at the microscopic level while still displaying collective
ordering transitions.

Recently, attention has shifted toward systems composed of two coupled
populations interacting through \emph{non-reciprocal} couplings~\cite{fruchart2021non,fruchart2026nonreciprocal,you2020nonreciprocity,saha2020scalar,scheibner2020odd,dinelli2023non,ivlev2015statistical,young2020nonequilibrium,avni2025nonreciprocal,avni2025dynamical,blom2025local,arjun2026kinetic}. 

In the simplest realization~\cite{avni2025nonreciprocal}, spins of species $A$ tend to align with spins of
species $B$, whereas spins of species $B$ preferentially anti-align
with those of species $A$. Such asymmetric interactions explicitly
break detailed balance and generate genuinely nonequilibrium dynamics,
opening the possibility of oscillatory phases and Hopf bifurcations that
have no equilibrium counterpart.

Such chase–avoid picture is particularly relevant to population dynamics, where nonreciprocal interactions between distinct populations $A$ and $B$ arise naturally in a variety of settings, 
including leader--follower dynamics~\cite{pomerenk2026minimal}, innovators and imitators~\cite{caulkins2007explaining}, cultural elites and mainstream populations~\cite{pedone2001dynamics}, or trend setters and consumers in fashion and financial markets~\cite{sornette2022non}. In all these cases, one population continuously adapts to the other, while the latter changes its state in response to being imitated, providing a generic microscopic mechanism for sustained collective oscillations.

Previous studies implemented non-reciprocity by directly modifying the local
field entering a Glauber transition rate~\cite{avni2025nonreciprocal,avni2025dynamical,arjun2026kinetic},
\begin{align}
W_K(s_i^X\rightarrow -s_i^X)
&=
\frac12
\bigg[
1-
s_i^X
\tanh
\bigg(
\beta H_i^X+ \nonumber \\
&\quad+K\sum_Y A_{XY}s_i^Y
\bigg)
\bigg]\,,
\label{WK}
\end{align}
where $\beta$ is the inverse temperature, $H_i=-\partial {\mathcal H}/{\partial s_i }$ is the local field, $K$ is the amplitude of the non-reciprocal coupling, $A_{XY}$ is an antisymmetric coupling matrix, and $X,Y \in \{A,B\}$, with $X\ne Y$ label the two species. 
In the \(q\)-voter model for sociodynamic systems, detailed balance cannot be invoked even at the level of an isolated subsystem. Instead, the dynamics is specified directly through a microscopic rule whereby a spin aligns with a group of $q$ randomly chosen spins in the system  with probability $W_0=1$, only if these $q$ spins are all oriented in the same direction. A temperature can be introduced by assuming $W_0<1$, but here we always consider zero temperature limit. Within this context it appears  more natural to implement non-reciprocal updates in an additive structure in terms of two competing stochastic processes,
\begin{align}
W(s_i^X\rightarrow -s_i^X)&=
(1-p)W_0(s_i^X\rightarrow -s_i^X)
\nonumber \\
&\quad+
\frac{p}{2}
\left(
1-\sum_Y A_{XY}s_i^Xs_i^Y
\right)\,,
\label{eq:additive}
\end{align}
where $W_0$ denotes the transition rate of the uncoupled model, i.e.,
Eq.~(\ref{WK}) for $K=0$ or the q-voter transition rate, while the second term represents a purely
non-reciprocal interaction occurring with probability $p$.
The coupling matrix
\begin{equation}
A=
\begin{pmatrix}
-\eta & 1+\eta\\
-(1-\eta) & \eta
\end{pmatrix}\,,
\label{eq:A}
\end{equation}
with $\eta\in[-1,1]$, is antisymmetric for $\eta=0$, whereas the parameter $\eta$ controls the
degree of non-normality  of the mutual interaction. Here $p\in[0,1]$ directly represents the probability (or frequency) of a
non-reciprocal interaction.

In this formulation, we observe that the non-reciprocal coupling matrix $A_{XY}$ appears linearly in the transition rate, whereas it enters via a hyperbolic tangent in Eq.~(\ref{WK}). In this Letter, we adopt the linear formulation which simplifies the analytical treatment, allowing us to uncover a remarkably rich nonequilibrium phenomenology. In particular, it enables us to focus directly on the zero-temperature limit, where the uncoupled system remains in its ordered or absorbing regime, and investigate the role of the remaining model parameters.
\begin{figure}[htb]
    
    \includegraphics[width=0.5\textwidth]{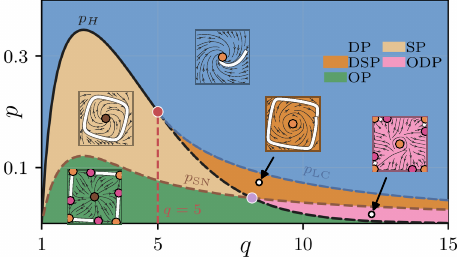}
    
    \caption{Phase diagram in the $(p,q)$ parameter space of the mean field uniformity $q$-voter model. The brown dashed line represents the saddle-node threshold, the black line represents the Hopf boundary (continuous and dashed indicate respectively a supercritical and a subcritical bifurcation), while the blue is the fold-limit-cycle curve. 
    For each dynamical phase an inset shows representative velocity fields in the \((m_A,m_B)\) plane, together with trajectories (white curves) originating from different initial condition. Orange, brown, and magenta dots denote stable, unstable, and saddle fixed points, respectively.}
    \label{fig:mf_pd}
\end{figure}
Introducing the magnetizations
\[
m_X=\langle s_i^X\rangle,\qquad X=A,B,
\]
where brackets indicates the average over configurations, the evolution equations become~\cite{glauber1963time}
\begin{equation}
    \dot m_X=(1-p)\big\langle F([s_i^X])\big\rangle-pm_X+p\sum_Y A_{XY}m_Y\,,
    \label{eq. general evolution}
\end{equation}
where $[s_i^X]$ denotes the microscopic spin configuration of subsystem $X$ and
\begin{equation}
    F([s_i^X])=-2\sum_{s_i^X=\pm1}s_i^XW_0(s_i^X\rightarrow-s_i^X)\,.
\end{equation}

Within mean-field theory, $F$ depends only on the mean magnetization, and Eq.~(\ref{eq. general evolution}) leads to the coupled pair 
\begin{equation}
\begin{aligned}
    \dot{m}_A &= (1-p)F(m_A) - pm_A + p \sum_X A_{AX} m_X\,, \\
    \dot{m}_B &= (1-p)F(m_B) - pm_B + p \sum_X A_{BX} m_X\,.
\end{aligned}
\label{eq. mean-field evolution}
\end{equation}
Eq.~(\ref{eq. mean-field evolution}) explicitly disentangles the structural properties inherited from the isolated dynamics (captured in $F$) from those driven by the non-reciprocal coupling $A_{XY}$.


Spin-inversion symmetry guarantees that the disordered configuration $(m_A,m_B)=(0,0)$ is always a fixed point. Near this disordered state, we expand the isolated response function as $F(m_X)=a(q)m_X+b(q)m_X^3+c(q)m_X^5+\mathcal{O}\left(m_X^7\right)$, where in all cases $c(q)$ is negative and we can neglect higher-order terms. Here, the coefficients $a, b,$ and $c$ explicitly depend on a control parameter $q$. In the $q$-voter model, $q$ retains its original microscopic meaning, namely the number of spins sampled in a unanimity update. In the following, however, we use $q$ in a broader sense to denote a generic model-dependent parameter that controls the critical behavior of the uncoupled system. For example, $q$ represents the amplitude of the external magnetic field in the RFIM~\cite{SM}.

The linear stability of the disordered fixed point is governed by the Jacobian eigenvalues,
\begin{equation}
    \lambda_\pm=\mu\pm i\Omega\,,\qquad\mu=(1-p)a(q)-p\,,\qquad \Omega=p\sqrt{1-2\eta^2}\,,
    \label{eq. eigenvalues}
\end{equation}
which form a complex-conjugate pair for all $|\eta|<1/\sqrt{2}$. The state loses stability when $\mathrm{Re}(\lambda_\pm) = 0$, marking the Hopf boundary~\cite{kuznetsov1998elements,strogatz2024nonlinear}
\begin{equation}
    p_H(q)=\frac{a(q)}{1+a(q)}\,.
    \label{eq. hopf boundary}
\end{equation}
The origin is stable for $p>p_H(q)$ and becomes unstable upon crossing $p_H(q)$ from above. Remarkably, $p_H(q)$ is completely independent of $\eta$: while non-normality affects the oscillation frequency, the eigenvectors, and the geometry of transient trajectories, it does not modify the asymptotic growth rate of the critical mode. Nevertheless, it can induce transient amplification even in the asymptotically stable regime~\cite{fruchart2026nonreciprocal}. The boundary of this regime, \(p_T(q,\eta)\), is defined by the vanishing of the numerical abscissa. Consequently, for \(p_H<p<p_T\), the origin remains asymptotically stable, while finite perturbations may undergo transient growth before eventually decaying.
Throughout the domain $|\eta|<1/\sqrt{2}$, branches of stationary ordered solutions cannot bifurcate continuously from the origin. Instead, static ordering can only emerge discontinuously at finite magnetization, controlled by the complete nonlinear response $F$.

For the normal case, $\eta=0$, the system exhibits $C_4$ symmetry, under which the nonzero solutions $(m_A,m_B)\neq(0,0)$ are invariant as a set under the transformation $(m_A,m_B)\rightarrow(m_B,-m_A)$. For $\eta\neq0$, this fourfold symmetry is explicitly reduced to global inversion. Although both families of ordered solutions, characterized by $m_A m_B>0$ and $m_A m_B<0$, continue to exist for all values of $\eta$, they are no longer equivalent: the sign of $\eta$ biases their relative stability, favoring the $m_A m_B>0$ family for $\eta>0$ and the $m_A m_B<0$ family for $\eta<0$.
The stability boundaries of the two branches define the saddle-node thresholds
$p_{\mathrm{SN}}^{(+)}(q,\eta)$ and $p_{\mathrm{SN}}^{(-)}(q,\eta)$,
associated with the branches $m_A m_B>0$ and $m_A m_B<0$, respectively.
Each branch remains stable for
$p<p_{\mathrm{SN}}^{(\pm)}(q,\eta)$,
where the superscript identifies the corresponding branch.
By symmetry, $p_{\text{SN}}^{(+)}(q,0)=p_{\text{SN}}^{(-)}(q,0) = p_{SN}(q)$.

For $|\eta|>1/\sqrt{2}$ the eigenvalues in Eq.~(\ref{eq. eigenvalues}) are real and the disordered state loses stability at $p=p_R(q,\eta)$. When $p<p_R$, ordered state may emerge directly from the disordered phase through a pitchfork bifurcation~\cite{kuznetsov1998elements,strogatz2024nonlinear,guckenheimer2013nonlinear}.

Following this general pattern, a remarkably rich phase diagram emerges even in the normal limit $\eta=0$. In Fig.~\ref{fig:mf_pd}, we present the phase diagram of the  $q$-voter model in the $(q,p)$ parameter space. Varying $q$ modulates both the Hopf boundary $p_H(q)$ and the saddle-node threshold $p_{\text{SN}}(q)$. Whenever $p_H(q) > p_{\text{SN}}(q)$, a dynamic swap phase (SP) exists for $p \in [p_{\text{SN}}(q), p_H(q)]$. Under these conditions, starting at $p=0$, the system begins in a stable ordered phase (OP)  configuration. As $p$ increases beyond $p_{\text{SN}}(q)$, the static state loses stability through a saddle-node on invariant cycle (SNIC) bifurcation, and the system enters the dynamic swap phase, where the two populations endlessly chase each other, until $p > p_H(q)$, where fluctuations dominate and the disordered phase (DP) is restored.

While this swap phase has been well documented in the context of the Ising model and the RFIM~\cite{avni2025nonreciprocal,avni2025dynamical,blom2025local,arjun2026kinetic}, a previously unreported regime emerges for $q > q^*$, where the two stability lines cross $p_H(q^*) = p_{\text{SN}}(q^*)$. Above this crossing (Fig.~\ref{fig:mf_pd}), $p_H(q) < p_{\text{SN}}(q)$, which precludes the swap phase; instead, both the ordered and disordered configurations are simultaneously stable and coexist (ODP). Consequently, a bistable region opens for $p \in [p_H(q), p_{\text{SN}}(q)]$, that can drive hysteresis~\cite{SM}.

The phase diagram is further enriched by a tricritical point~\cite{kuznetsov1998elements,arjun2026kinetic,collet2019effects} at $q=5$, where the cubic coefficient $b(q)$ vanishes and becomes positive for $q>5$. This switches the Hopf bifurcation from supercritical to subcritical, giving rise to a new transition boundary $p_{LC}(q)$ that originates at $p_H(q=5)$~\cite{kuznetsov1998elements,guckenheimer2013nonlinear} giving rise to a fold-of-limit-cycles curve $p_{\rm LC}(q)$ emanating from the generalized Hopf point $(q,p)=(5,p_H(5))$. In this case, crossing $p_{H}(q)$ drives a discontinuous transition from the disordered phase to finite-amplitude oscillations. In the subcritical regime between $p_H(q)$ and $p_{LC}(q)$, the linearly stable origin coexists with a stable limit cycle (DSP), enabling hard-mode excitations and hysteresis~\cite{kuznetsov1998elements,guckenheimer2013nonlinear,goldbeter2002computational,epstein1996nonlinear,subramanian2013subcritical,armen2006low}.

A detailed analytical derivation of the complete phase diagram and further characterization of the individual phases are provided in~\cite{SM}. Crucially, although demonstrated here for the $q$-voter model, this local phase structure is entirely generic, relying only on the functional dependence of the coefficients $a,b,$ and $c$ on the control parameters. For example, the full phenomenology is reproduced in the RFIM, where $q$ parameterizes the strength of the random field~\cite{SM}.

A specificity of the $q$-voter model compared to the RFIM is the presence of a reentrant transition. Indeed, at an appropriate range of fixed $p$, increasing $q$ from one, causes the system to transition from the disordered phase into the swap phase, and then reenter the disordered phase as $q$ continues to grow.

\begin{figure}[htb]
    
    \includegraphics[width=0.5\textwidth]{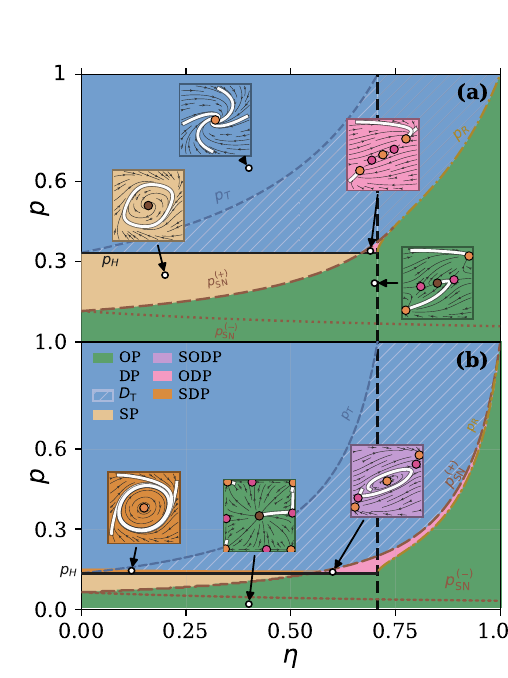}
    
    \caption{Phase diagram in the $(p,\eta)$ parameter space of the mean field non-normal uniformity $q$-voter model. Panel (a) refers to results for $q=2$, while Panel (b) to $q=6$. The brown dashed lines represent the saddle-node threshold, the black continuous line represents the Hopf boundary, the black dashed line is $\eta=1/\sqrt{2}$, while the blue is the fold-limit-cycle curve. 
    The insets show representative velocity fields and trajectories in
the $(m_A,m_B)$ plane, using the same graphical conventions as in
Fig.~\ref{fig:mf_pd}.}
    \label{fig:nonnormal_pd}
\end{figure}

The zero-temperature Ising model corresponds to the $q=2$ limit of the $q$-voter model. Remarkably, even in the baseline case $q=2$, non-normality induces bistability and substantially enriches the phase behavior. Indeed non-normality breaks the $C_4$ symmetry, leaving only two stable ordered fixed points in each ordered regime and deforms the periodic orbits into ellipses whose principal axes are no longer aligned with the $m_A$ and $m_B$ coordinate axes. More notably, varying $\eta$ gives rise also for $q=2$ to the nontrivial OD phase, as shown in the phase diagram in Fig.~2(a). Indeed, as $\eta$ increases, $p_{\text{SN}}^{(+)}(2,\eta)$ increases while $p_{\text{SN}}^{(-)}(2,\eta)$ decreases. Since the Hopf boundary $p_H(2)$ is $\eta$-independent, a crossover threshold $\eta^*$ emerges where $p_{\text{SN}}^{(+)}(2,\eta^*)=p_H(2)$. For $\eta > \eta^*$, a bistable window opens for $p \in [p_H(2), p_{\text{SN}}^{(+)}(2,\eta)]$, characterized by the coexistence of the disordered state and the aligned ordered phase. Non-normality thus introduces bistability into an otherwise simpler scenario. Fig.~\ref{fig:nonnormal_pd}(a) depicts this scenario for $\eta \ge 0$. An analogous picture holds for $\eta < 0$, where decreasing $\eta$, $p_{\text{SN}}^{(-)}(2,\eta)$ grows while $p_{\text{SN}}^{(+)}(2,\eta)$ shrinks, yielding a bistable region $p \in [p_H(2), p_{\text{SN}}^{(-)}(2,\eta)]$ where the disordered state coexists with an anti-aligned ordered phase ($m_A m_B < 0$).
Introducing non-normality $\eta$ in the regime beyond the tricritical point ($q > 5$) yields an even richer phase structure. In particular, we uncover a narrow tristable window where three distinct attractors—a stable limit cycle, an ordered fixed point, and the disordered state (SODP)—simultaneously coexist in a narrow tristable region, denoted SODP in Fig.~\ref{fig:nonnormal_pd}(b) for $q=6$.

At $|\eta|=1/\sqrt 2$ and $p>0$, the Jacobian becomes defective, with coalescing eigenvalues and eigenvectors, defining two exceptional lines~\cite{weis2025generalized,zelle2024universal,kuznetsov1998elements,hanai2020critical,guckenheimer2013nonlinear}. Their intersections with the Hopf boundary $p=p_H(q)$ define two critical exceptional points~\cite{bogdanov1975versal,bogdanov1976bifurcations,takens2001forced}. Details can be found in~\cite{SM}

\textit{Finite dimensions.} We now turn to consider the case of finite dimensions, close to the Hopf line (Eq.~\ref{eq. hopf boundary}) and in the complex-eigenvalues sector. After coarse-graining and a gradient expansion, the deterministic part of the dynamics takes the Landau--Ginzburg form
\begin{equation}
F[\phi]=a(q)\phi+D\nabla^2\phi+b(q)\phi^3+c(q)\phi^5+\mathcal O(\phi^7)\,.
\end{equation}
The coupled equations for the field $\phi_X$ then read 

$$
\begin{aligned}
\partial_t\phi_A &= \left[(1-p)a(q)-p(1+\eta)\right]\phi_A+p(1+\eta)\phi_B+ \\
&\quad+(1-p)D\nabla^2\phi_A+(1-p)b(q)\phi_A^3+(1-p)c(q)\phi_A^5\,,
\\
\partial_t\phi_B &= -p(1-\eta)\phi_A+\left[(1-p)a(q)-p(1-\eta)\right]\phi_B+ \\
&\quad+(1-p)D\nabla^2\phi_B+(1-p)b(q)\phi_B^3+(1-p)c(q)\phi^5_B\,.
\end{aligned}
$$
The coupled fields inherit the  discrete $\mathbb{Z}_2$ symmetry. Introducing the complex field
\begin{equation}
    \Psi=\frac{\sqrt{1-2\eta^2}-i\eta}{\sqrt{1+\eta}}\phi_A+i\sqrt{1+\eta}\phi_B\,,
\end{equation}
and projecting the dynamics onto the critical eigenspace generates both the isotropic nonlinearities $|\Psi|^2\Psi$ and $|\Psi|^4\Psi$, and anisotropic terms such as $(\Psi^*)^3$. The latter are non-resonant, since they oscillate at frequency $-3\Omega$, whereas the critical mode evolves at frequency $\Omega$. They can be eliminated through a near-identity transformation. The same mechanism removes all higher-order anisotropic contributions, leaving only the resonant nonlinearities. Details are presented in~\cite{SM}. Consequently, independently of the microscopic realization, the long-wavelength critical dynamics is governed by the quintic complex Ginzburg--Landau equation~\cite{aranson2002world}
\begin{equation}
\partial_t\Psi=(\mu-i\Omega)\Psi+D\nabla^2\Psi
+g_3(q,\eta)|\Psi|^2\Psi
+g_5(q,\eta)|\Psi|^4\Psi\,,
\label{eqXY}
\end{equation}
where $\mu$ and $\Omega$ were defined in Eq.~(\ref{eq. eigenvalues}), while $g_3(q,\eta)$ and $g_5(q,\eta)$ have a nontrivial dependence on the expansion parameters~\cite{SM}. Remarkably, the continuous $U(1)$ phase symmetry of the amplitude equation emerges dynamically from the underlying discrete $\mathbb{Z}_2$ symmetry in the resonant normal form near the Hopf instability.

If we consider the situation where the quintic term is irrelevant as in the case of  Ising model~\cite{avni2025nonreciprocal,avni2025dynamical} or the $q$-voter model for $q<5$, the rotating-frame dynamics reduces to a purely relaxational O(2)-symmetric time-dependent Ginzburg-Landau equation. Therefore, the stationary critical properties belong to the XY universality class~\cite{risler2004universal,risler2005universal}.
Accordingly, in two
dimensions, the continuous symmetry prevents conventional long-range
order~\cite{mermin1966absence} and the transition, if present, is expected to be of
Berezinskii--Kosterlitz--Thouless type~\cite{berezinskii1971destruction,kosterlitz1973ordering}. Conversely, for $d\geq3$ the
system can display a genuine order-disorder transition belonging to the
$O(2)$ universality class.
This scenario is supported by numerical simulations~\cite{avni2025nonreciprocal,avni2025dynamical}, which show that coherent oscillatory order is suppressed in $d=2$, consistently with the absence of long-range $O(2)$ order, whereas stable oscillations survive in $d=3$. In the latter case, finite-size scaling yields critical exponents compatible with the $O(2)$ (XY) universality class~\cite{avni2025nonreciprocal,avni2025dynamical}.

Writing $\Psi=\rho e^{i\theta}$, where $\rho$ and $\theta$ denote the local amplitude and phase, respectively, we adiabatically eliminate the rapidly relaxing amplitude by setting it to its local quasi-stationary value, which adjusts instantaneously to the slowly varying phase gradients. The resulting leading-order phase dynamics is
\begin{equation}
\partial_t\theta=\nu\nabla^2\theta+\gamma(\nabla\theta)^2,
\label{eq_theta}
\end{equation}
where $\nu$ is the effective phase diffusivity and $\gamma\propto\operatorname{Im}g_3(q,\eta)$ arises from nonlinear amplitude--phase coupling. Accordingly, in the normal case $\eta=0$, an important difference emerges when comparing the dynamics
with linear non-reciprocity  (Eq.~(\ref{eq:additive})) in the supercritcal regime, considered here, and the hyperbolic one (Eq.~(\ref{WK})) considered in~\cite{avni2025nonreciprocal,avni2025dynamical}. In both situations, the coarse-grained description
leads to a complex Ginzburg--Landau equation of the form
Eq.~(\ref{eqXY}), truncated at the cubic term.
However, in the linear case $g_3(q)$ is real whereas in the hyperbolic case $g_3(q)$ acquires an imaginary component. 
Consequently, a linear non-reciprocity leads to an Edwards--Wilkinson phase
equation ($\gamma=0$)~\cite{edwards1982surface} in Eq.~(\ref{eq_theta}), whereas hyperbolic non-reciprocity generates a
genuine Kardar--Parisi--Zhang (KPZ) dynamics ($\gamma\neq0$)~\cite{kardar1986dynamic}.

\begin{figure}[htb]
    
    \includegraphics[width=0.5\textwidth]{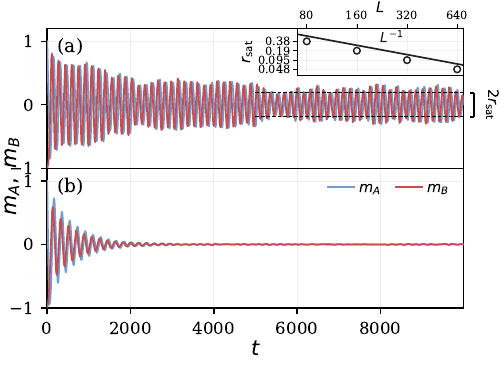}
    
    \caption{Magnetizations $m_A$ (blue curve) and $m_B$ (red curve) as functions of time $t$ in the two-dimensional $q$-voter model with $q=2$, $p=0.07$, and $L=160$. Panel (a) shows the normal case, $\eta=0$, while panel (b) shows the non-normal case, $\eta=0.2$. In the normal case, the system relaxes toward a limit cycle with a finite saturation radius $r_{\mathrm{sat}}$, that oscillates over time. The inset shows the log-log plot of the finite-size scaling $r_{\mathrm{sat}}\propto L^{-1}$, obtained for system sizes $L=80,160,320,640$.}
    \label{fig:mag_vs_time}
\end{figure}

This distinction has consequences for the stability of the
oscillatory phase in two dimensions. The KPZ nonlinearity induces a
nonlinear phase advection that destabilizes long-range phase coherence,
leading to the proliferation of spiral-like defects observed in
simulations of hyperbolic non-reciprocal systems~\cite{avni2025nonreciprocal,avni2025dynamical}. Conversely, in the
linear non-reciprocal case no such defect-mediated instability is
observed, and coherent oscillations persist in finite systems.
This behavior is illustrated in Fig.~\ref{fig:mag_vs_time}(a), where we plot the magnetization dynamics obtained from numerical simulations of the $q$-voter model $(q=2)$ with linear non-reciprocity (Eq.~(\ref{eq:additive})) and $\eta=0$.  
We observe that oscillations persist and approach a finite-size saturation amplitude $r_{\mathrm{sat}}$. Nevertheless, this oscillatory order disappears in the thermodynamic limit: as shown in the inset of Fig.~\ref{fig:mag_vs_time}, $r_{\mathrm{sat}}$ decreases with system size as $r_{\mathrm{sat}}\sim L^{-1}$. This scaling behavior admits a physical interpretation in terms of the nonequilibrium dynamics of topological defects~\cite{SM}.

Conversely, even in the case of a linear non-reciprocity as in Eq.~(\ref{eq:additive}), non-normality ($\eta\neq0$) induces an imaginary part in $g_3(q)$, thereby leading to a KPZ-type description of the phase dynamics. In this case, the oscillations are rapidly suppressed, as shown in Fig~\ref{fig:mag_vs_time}(b).
    
In conclusions, our results indicate that the nonequilibrium behavior induced by nonreciprocal coupling is shaped by the critical properties of the isolated systems. This connection may extend beyond the models considered here and suggests that coupling multiple critical subsystems could generate richer collective modes. More broadly,  nonreciprocal coupling may provide a general mechanism for transferring, reshaping, and combining critical behavior across interacting systems.

\bibliographystyle{apsrev4-2}
\bibliography{references}

\end{document}